# Classification of Imagined Speech Using Siamese Neural Network

Dong-Yeon Lee
*Dept. Brain and Cognitive Engineering*
*Korea University*
Seoul, Republic of Korea
dongyeon_lee@korea.ac.kr

Minji Lee
*Dept. Brain and Cognitive Engineering*
*Korea University*
Seoul, Republic of Korea
minjilee@korea.ac.kr

Seong-Whan Lee
*Dept. Artificial Intelligence*
*Dept. Brain and Cognitive Engineering*
*Korea University*
Seoul, Republic of Korea
sw.lee@korea.ac.kr

*Abstract*— Imagined speech is spotlighted as a new trend in the brain-machine interface due to its application as an intuitive communication tool. However, previous studies have shown low classification performance, therefore its use in real-life is not feasible. In addition, no suitable method to analyze it has been found. Recently, deep learning algorithms have been applied to this paradigm. However, due to the small amount of data, the increase in classification performance is limited. To tackle these issues, in this study, we proposed an end-to-end framework using Siamese neural network encoder, which learns the discriminant features by considering the distance between classes. The imagined words (e.g., arriba (up), abajo (down), derecha (right), izquierda (left), adelante (forward), and atrás (backward)) were classified using the raw electro-encephalography (EEG) signals. We obtained a 6-class classification accuracy of 31.40 ± 2.73% for imagined speech, which significantly outperformed other methods. This was possible because the Siamese neural network, which increases the distance between dissimilar samples while decreasing the distance between similar samples, was used. In this regard, our method can learn discriminant features from a small dataset. The proposed framework would help to increase the classification performance of imagined speech for a small amount of data and implement an intuitive communication system.

## I. Introduction

Brain-machine interface (BMI) has multiple applications such as control external devices, communication systems, and more; for which it uses brain signals. Traditionally this interface has been used by severely disabled people, however, recently it is being used by healthy people for communication or daily life assistance systems [1-3]. Among them, electroencephalography (EEG) is commonly used due to its low cost and convenience in the field of BMI [4, 5]. Under the BMI paradigm, there are two ways to generate brain signals [6]. One is exogenous paradigms and the other is endogenous paradigms. Exogenous paradigms require the presentation of an external stimulus such as flickering visual stimulation, auditory stimulation, and more to evoked brain signals [7-9]. On the other hand, endogenous paradigms do not require an external stimulus [10]; subjects perform a mental task (i.e., motor imagery (MI), visual imagery, and imagined speech) while brain signals are recorded [11, 12]. Endogenous paradigm can reflect the user's intention and be applied to more intuitive systems, therefore, its application in BMI technology has been increasing.

The most widely used paradigm in the conventional endogenous BMI is MI. In MI paradigms, brain signals are generated through the imagination of movement intention [10]. It is commonly used to control prosthetic devices such as robot arms. However, the number of imaginable MI classes is limited (i.e., left hand, right hand, tongue, and foot). When using MI for communication systems such as spellers [13], it cannot be an intuitive system; because the user imagines not the word they want to enter but a specific action [14, 15]. Therefore, an intuitive paradigm to solve these problems has begun to come into the spotlight.

Imagined speech is a new trend of BMI paradigm that conduct internal pronunciation of words without any movement and audible output [16]. This paradigm until now has employed traditional BMI feature extraction and classification method [14, 15, 17]. The commonly used feature extraction method is common spatial pattern (CSP) [14, 17]. In Lee et al. [14], they classified 13-classes of imagined speech using the CSP as a feature extraction method. For classification methods, shrinkage regularized linear discriminant analysis (RLDA) and random forest (RF) were used. As a result, RF showed the highest accuracy; 20.4% for 13-class classification (12 imagined words with the resting state). In Dasalla et al. [17], CSP and support vector machine (SVM) were applied for binary vowel imagery classification. The overall classification accuracies ranged from 68% to 78%. CSP has the advantage to reduce the high dimensional signal into the low dimensional signal and can maximize the distance between different classes. However, CSP is optimized for binary class problems [18]. Therefore, other methods were also used for feature extraction. In Coretto et al.

This work was supported in part by the Institute for Information & Communications Technology Promotion (IITP) grant, funded by the Korea government (MSIT) (No. 2015-0-00185, Development of Intelligent Pattern Recognition Softwares for Ambulatory Brain Computer Interface; No. 2017-0-00451, Development of BCI based Brain and Cognitive Computing Technology for Recognizing User's Intentions using Deep Learning; No. 2019-0-00079, Artificial Intelligence Graduate School Program(Korea University))

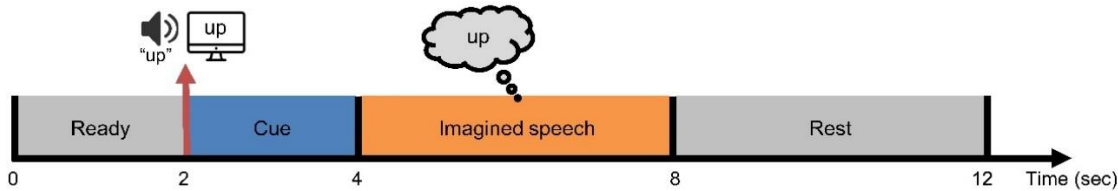

Fig. 1. Experimental procedure of imagined speech.

[19], classification of 5 vowels and 6 words was aimed, they used discrete wavelet transform (DWT) for feature extraction and RF for classification. These methods showed performances close to the chance level (vowel: 22.72%, words: 19.60%). In García-Salinas et al. [20], they used a bag of features (BoF) for feature extraction and Naive Bayes for classification. They showed an average accuracy of 65.65% in 5-class imagined word classification. However, they claimed that the performance was high because the presentation of stimuli during the data collection process was not randomized. Until now, these feature extraction and classification method have not yet shown optimal performance in the imagined speech paradigm [21]. Thus, imagined speech classification performance can be improved by selecting appropriate features and classifiers [22]. However, it is not easy to find a suitable analysis method for imagined speech.

To tackle these issues and improve performance, recently, studies on analyzing imagined speech through deep learning have been increasing. A commonly used deep learning method is convolutional neural network (CNN). In Saha et al. [23], to classify binary classification of phonological categories, they employed the channel cross-covariance for preprocessing. The model framework was composed of CNN, long-short term network (LSTM), and a deep autoencoder to extract the spatio-temporal information. They achieved the best average accuracy of 77.9% across five different binary classification tasks. In their recent work [24], they employed channel cross-covariance for preprocessing and composed the network of spatial and temporal CNN cascaded with a deep autoencoder to perform binary classification of phonological categories. Obtained average accuracy was 83.42% across the six different binary phonological classifications. In Cooney et al. [21], they employed independent component analysis (ICA) with Hessian approximation preconditioning and CNN to classify 5 imagined vowels and an accuracy of 32.35% was obtained. In their recent work using ICA with Hessian approximation preconditioning, DeepConvNet [25] and ShallowConvNet [25] were applied [26]. They achieved 62.37% accuracy for word-pairs classification. Although various deep learning methods have been applied, they have not yet been able to perform as well as in endogenous BMI paradigm such as MI.

Usually, DeepConvNet [25] and ShallowConvNet [25] are commonly used for raw EEG signal analysis. However, in Cooney et al. [26], they used additionally preprocessing or feature extraction methods. Few studies use end-to-end learning for imagined speech [27]. Preprocessing or feature extraction can lead to loss of information. These procedures also require a computational cost. Therefore, the end-to-end learning method with minimal preprocessing or feature extraction is increasing interest in EEG classification. In addition, one of the important issues in EEG classification is the lack of trials and high dimensionality [28]. Therefore, finding a suitable deep learning method for EEG classification using raw EEG signals is also emerging as a new interest in the field of BMI.

To solve the aforementioned problems, we used the Siamese neural network encoder. The Siamese neural network takes two inputs and trains the same network in parallel [28]. It aims to optimize the distance in the embedding space. For example, if two randomly selected samples are of the same class, they are learned to be located close to each other. And if they are different classes, they are learned to be located far from each other. The Euclidean distance is used to measure the distance of the embeddings extracted from the Siamese neural network [28].

In this study, we proposed an end-to-end Siamese neural network encoder-based classification method. To best of our knowledge, there is no approach to classify the imagined speech using Siamese neural network encoder. Dataset consists of 6 words, based on EEG signals for the imagined speech. The embeddings are extracted based on the Siamese neural network and $k$-nearest neighbor ($k$-NN) was used for classification. This approach can extract discriminant features corresponding to each class from raw data. Therefore, the proposed method can increase the performance of imagined speech for its applications in real-life environments.

## II. MATERIAL AND METHODS

### A. Data Description

The dataset used for this study is the open data of Coretto et al [19]. It consists of EEG signals from 15 subjects. EEG signals were recorded using Ag-AgCl electrodes. The electrodes were placed over F3, F4, C3, C4, P3, and P4 while reference and ground electrodes were placed over left and right mastoids, respectively. These electrodes partially included the Wernicke area related to language processing [15, 22]. Subjects performed the imagined speech and pronounced speech of 5 Spanish vowels and 6 Spanish words. In this study, we use only imagined speech EEG signals corresponding to 6 words. The number of imagined speech trials performed by each subject was 40 per word. The command words were arriba (up), abajo (down), derecha (right), izquierda (left), adelante (forward), and atrás (backward). These commands were chosen to intuitively control external devices in a BMI system.

The experiment paradigm consists of 4 procedures per trial as shown in Fig. 1. In each trial, a ready interval is presented first for 2 sec. In this procedure, the subjects are instructed to relax. Then visual, and audible cues are presented for 2 sec. The visual cue is presented on the monitor and audible cue is given through speakers. When the cue disappeared, the subjects are instructed to imagine the pronunciation of the word presented by the cue for 4 sec. The final procedure is a rest period for 4 sec. These four procedures compose one trial, and this trial is repeated 40 times per each word randomly.

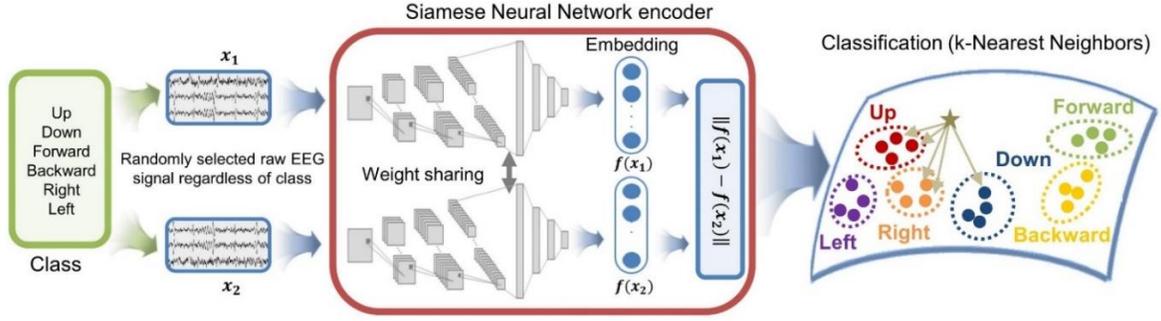

Fig. 2. Proposed framework of end-to-end Siamese neural network encoder-based classification. One trial in the test dataset is marked with a star. When the test trial is located in the embedding space via the Siamese neural network, the class is determined based on the class of five nearest instances.

TABLE I. NEURAL ARCHITECTURE FOR SIAMESE NEURAL NETWORK

| Layer | Input | Output | Kernel |
|---|---|---|---|
| Conv1 | $6 \times 512 \times 1$ | $2 \times 103 \times 64$ | $5 \times 5$ |
| Maxpool1 | $2 \times 103 \times 64$ | $1 \times 52 \times 64$ | $2 \times 2$ |
| Conv2 | $1 \times 52 \times 64$ | $1 \times 52 \times 128$ | $1 \times 3$ |
| Maxpool2 | $1 \times 52 \times 128$ | $1 \times 52 \times 128$ | $1 \times 3$ |
| Conv3 | $1 \times 52 \times 128$ | $1 \times 52 \times 128$ | $1 \times 2$ |
| Maxpool3 | $1 \times 52 \times 128$ | $1 \times 52 \times 128$ | $1 \times 2$ |
| Flatten | $1 \times 52 \times 128$ | 6656 | - |
| FC1 | 6656 | 1024 | - |
| FC2 | 1024 | 512 | - |
| FC3 | 512 | 256 | - |
| FC4 | 256 | 8 | - |

EEG signals are sampled at 1,024 Hz, this high dimensional signal is down-sampled to 128 Hz. Down-sampling aims to reduce the amount of data [19]. The original dataset was band-pass filtered between 2 and 40 Hz using a finite impulse response (FIR) band-pass filter. In this study, no further processing, such as preprocessing or feature extraction, was performed.

*B. Proposed Method*

*1) Overall framework:* The overall flow chart is shown in Fig. 2. First, raw data is input to the Siamese neural network. Each data consists of 4sec and 6 channels. Two inputs are required to be used on the Siamese neural network. Therefore, two random samples are selected from the total data. These two samples are selected regardless of the class. Each sample trains a CNN branch, both branches have the same structure and parameters. If the classes of the two data are the same, the embedding extracted through the Siamese neural network is learned to be close to each other, and in case the classes of the two data are different, the embedding is learned to be distant. As a result, the Siamese neural network learns the distances between samples in the specific space. By using Siamese neural networks over high-dimensional raw data, we can get dimensionally reduced embedding that contains important information related to the distance between different classes. For classification, k-NN was used; it predicts the classes of test data sets through the closest k instances in a specific space.

*2) Feature extraction using Siamese neural network:* We used the Siamese neural network to address the drawback of a small amount of EEG trial. Because of the small number of trials in EEG data, using conventional deep learning methods are more likely to cause overfitting. On the other hand, the Siamese neural network belongs to metric learning. Siamese neural network learns from two inputs and extracts reduced embedding. The two networks are identical and share parameters. Inputs are randomly selected in the training set. These randomly selected inputs are denoted $x_1$, $x_2$. The extracted embeddings through the network are also denoted as $f(x_1), f(x_2)$. The network is trained to distinguish whether the two inputs are the same or not [18]. Therefore, y is a label that indicates whether the input is in the same class ($y = 1$) or different class ($y = 0$). In order to train the Siamese neural network, information related to labels needs to be given to the network in each iteration. The information structure is $[x_1, x_2, y]$ which indicates the two inputs and the corresponding label. When the embedding is extracted through the network, the Euclidian distance between these two embeddings is calculated as follows:

$$D(x_1, x_2) = \|f(x_1) - f(x_2)\|_2 \quad (1)$$

Many deep learning methods use cross-entropy loss. However, cross-entropy losses focus on predicting the class the sample belongs to. Therefore, it is not suitable for learning the distance between the embedding [29]. The Siamese neural network uses contrastive loss. Contrastive loss learns that samples belonging to the same class are placed close to each other, and samples belonging to different classes are placed far away in the embedding space [18, 28]. The contrastive loss function is as follows:

$$L(x_1, x_2, y) = \tfrac{1}{2} y D^2 + \tfrac{1}{2}(1-y)\max(m - D, 0)^2 \quad (2)$$

where *m* denotes a margin ($m > 0$). This is a parameter set by the user and plays a role in making the distance between two samples greater its value when two samples are from different classes [29]. The Siamese neural network was originally proposed for one-shot learning [30], in contrast, this study aims to increase the performance by learning as many combinations of inputs as possible in the training set. We use the embedding obtained from the Siamese neural network to train the classifier; our purpose is to classify samples in different classes, not to determine whether two classes are the same or different.

Our Siamese neural network architecture is specified in Table I. The encoder scheme was trained using the Siamese neural network. Siamese neural network was trained using the ADAM optimizer and the contrastive loss function. We set the gradient decay factor to 0.9 and the squared gradient decay factor to 0.99. We also set the margin m=0.5, the learning rate to $1 \times 10^{-4}$. The batch size is 180 for each iteration and train for 1,000 iterations. The Rectified linear unit (ReLU) was used after each convolution layer and fully-connected layer.

*3) Classification using k-NN:* From the Siamese neural network, we get a reduced dimension embedding from the high dimensional original data. The k-NN is used to predict classes using the extracted embedding [31]. The nearest number of neighbors is set to 5 in the k-NN method (k = 5). k-NN is used because it makes predictions based on the nearest k instance in the specific embedding space. The Siamese neural network also learns the distance between data in specific embedded spaces, so k-NN is more suitable for analysis than other classification methods.

### C. Training and Evaluation Scheme

To evaluate the classification performance, 5-fold cross-validation was performed for each subject. In each class, the same number of trials are used for training. Other methods are evaluated with the same 5-fold cross-validation approach for each subject. As classification measures, we computed precision, recall, F1-score, and accuracy for 6-class.

### D. Statistical Analysis

We performed a statistical analysis to estimate the significant difference between our proposed method and different methods. The one-way analysis of variance (ANOVA) was used, and post-hoc analysis was performed using paired t-test. For multiple comparisons, Bonferroni correction was applied. All significant values were set at *p*-value = 0.05.

## III. RESULTS

### A. Classification Performance using Proposed Method

Table II shows classification accuracy using the proposed method across all subjects. The 6-class classification using our proposed method shows an average accuracy of 31.40 ± 2.73 % across all subjects. Subject 4 showed the highest performance (36.51 ± 4.13%).

Fig. 3. illustrates the confusion matrix. It is observed that "up" shows the best classification performance. However, all words have similar performance. Furthermore, all obtained accuracies were higher than the chance level (16.67%). Table III shows recall and F1-score in all subjects. The class "forward" showed the highest performance, whereas "up" showed the lowest performance in both recall and F1-score.

TABLE II. 6-CLASS CLASSIFICATION ACCURACY USING PROPOSED METHOD ACROSS ALL SUBJECTS

| Subject | Accuracy (%) |
|---|---|
| Sub01 | 28.84 ± 2.51 |
| Sub02 | 34.43 ± 3.15 |
| Sub03 | 30.77 ± 1.90 |
| Sub04 | 36.51 ± 4.13 |
| Sub05 | 30.95 ± 1.06 |
| Sub06 | 33.93 ± 3.66 |
| Sub07 | 32.69 ± 2.51 |
| Sub08 | 31.64 ± 2.11 |
| Sub09 | 27.14 ± 1.20 |
| Sub10 | 33.82 ± 2.23 |
| Sub11 | 29.49 ± 1.67 |
| Sub12 | 29.16 ± 0.76 |
| Sub13 | 29.69 ± 1.71 |
| Sub14 | 27.38 ± 0.65 |
| Sub15 | 34.55 ± 3.04 |
| **Average ± Std.** | **31.40 ± 2.73** |

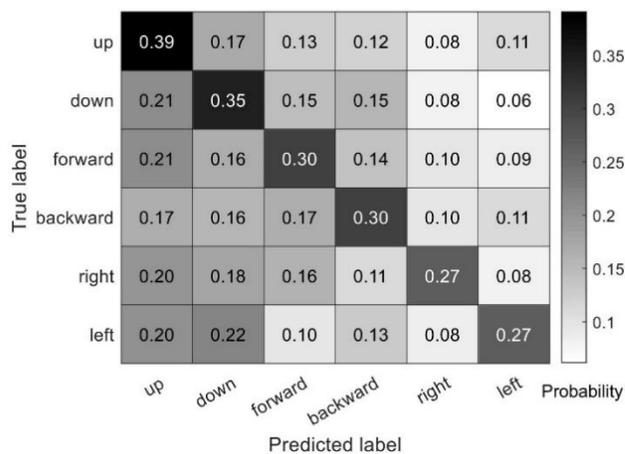

Fig. 3. Averaged confusion matrix using proposed method across all subjects.

### B. Comparison of Conventional Methods

We compared the classification performance of the proposed framework and conventional methods. Our method improved performance by 9.93 to 13.94% compared to the other baseline classification method. Table IV shows the averaged classification accuracy of different methods across all subjects and the results of the statistical analysis. There were significant difference between the accuracies of different methods ($F_{(4)}$ = 164.43, *p*-value < 0.001). Additionally, the post-hoc test revealed significant differences between the proposed method and other four conventional methods (*p*-value < 0.001).

## IV. DISCUSSION

The proposed end-to-end Siamese neural network encoder-based classification method shows that the Siamese neural network can learn discriminant features from raw data. In particular, it shows better performance than machine learning methods that use preprocessing or feature extraction. Traditional machine learning methods have many limitations when training models using raw data. Therefore, the preprocessing or feature extraction process is absolutely necessary before classification or clustering [29]. However, it is not easy to find a suitable preprocessing or feature extraction method. It is due to when the data type is different, the corresponding characteristics also change. The proposed method does not require a preprocessing or feature extraction procedure since Siamese neural network operates as a feature extractor. This is because deep learning can directly learn high dimensional raw data [29].

TABLE III. RECALL AND F1-MEASURE PER EACH IMAGINED SPEECH WORD (6-CLASS) ACROSS ALL SUBJECTS

| Subject | Up | | Down | | Right | | Left | | Forward | | Backward | |
|---|---|---|---|---|---|---|---|---|---|---|---|---|
| | Recall | F1-score | Recall | F1-score | Recall | F1-score | Recall | F1-score | Recall | F1-score | Recall | F1-score |
| Sub01 | 0.368 | 0.389 | 0.250 | 0.200 | 0.222 | 0.229 | 0.316 | 0.333 | 0.278 | 0.270 | 0.278 | 0.294 |
| Sub02 | 0.214 | 0.250 | 0.333 | 0.400 | 0.571 | 0.444 | 0.333 | 0.316 | 0.500 | 0.375 | 0.300 | 0.300 |
| Sub03 | 0.278 | 0.345 | 0.214 | 0.222 | 0.294 | 0.345 | 0.333 | 0.320 | 0.429 | 0.286 | 0.400 | 0.320 |
| Sub04 | 0.308 | 0.364 | 0.500 | 0.444 | 0.333 | 0.364 | 0.333 | 0.300 | 0.500 | 0.421 | 0.308 | 0.320 |
| Sub05 | 0.154 | 0.200 | 0.250 | 0.167 | 0.286 | 0.308 | 0.300 | 0.353 | 0.750 | 0.545 | 0.500 | 0.364 |
| Sub06 | 0.235 | 0.348 | 0.250 | 0.286 | 0.308 | 0.348 | 0.600 | 0.429 | 0.400 | 0.286 | 0.750 | 0.353 |
| Sub07 | 0.214 | 0.260 | 0.267 | 0.381 | 0.333 | 0.267 | 0.500 | 0.400 | 0.429 | 0.375 | 0.500 | 0.357 |
| Sub08 | 0.231 | 0.250 | 0.250 | 0.333 | 0.364 | 0.348 | 0.273 | 0.231 | 0.357 | 0.370 | 0.500 | 0.357 |
| Sub09 | 0.333 | 0.300 | 0.417 | 0.417 | 0.333 | 0.286 | 0.143 | 0.188 | 0.200 | 0.182 | 0.333 | 0.286 |
| Sub10 | 0.333 | 0.381 | 0.250 | 0.261 | 0.417 | 0.417 | 0.235 | 0.276 | 0.500 | 0.400 | 0.429 | 0.316 |
| Sub11 | 0.278 | 0.345 | 0.194 | 0.293 | 0.182 | 0.174 | 0.600 | 0.316 | 0.333 | 0.182 | 0.714 | 0.455 |
| Sub12 | 0.211 | 0.286 | 0.333 | 0.370 | 0.125 | 0.105 | 0.286 | 0.200 | 0.455 | 0.417 | 0.333 | 0.308 |
| Sub13 | 0.333 | 0.364 | 0.250 | 0.273 | 0.214 | 0.250 | 0.333 | 0.381 | 0.429 | 0.286 | 0.286 | 0.222 |
| Sub14 | 0.211 | 0.242 | 0.236 | 0.258 | 0.294 | 0.345 | 0.267 | 0.276 | 0.364 | 0.308 | 0.400 | 0.200 |
| Sub15 | 0.190 | 0.286 | 0.182 | 0.200 | 0.250 | 0.250 | 0.714 | 0.588 | 1.000 | 0.286 | 0.667 | 0.533 |
| **Average** | **0.259** | **0.307** | **0.278** | **0.300** | **0.302** | **0.298** | **0.371** | **0.327** | **0.461** | **0.333** | **0.446** | **0.328** |
| **Std.** | **0.062** | **0.057** | **0.082** | **0.084** | **0.102** | **0.087** | **0.153** | **0.096** | **0.187** | **0.093** | **0.152** | **0.079** |

TABLE IV. STATISTICAL RESULTS IN AVERAGE CLASSIFICATION PERFORMANCE COMPARED TO PROPOSED METHOD

| Method | Accuracy (%) | t-value | p-value |
|---|---|---|---|
| Coretto et al. [17] | 17.46 ± 0.71 | -18.81 | <0.001 |
| Cooney et al. [19] | 18.89 ± 1.41 | -15.08 | <0.001 |
| Schirrmeister et al. [23] | 19.81 ± 2.10 | -14.42 | <0.001 |
| García-Salinas et al. [18] | 21.47 ± 1.99 | -14.43 | <0.001 |
| Proposed method | 31.40 ± 2.73 | - | - |

Especially, our proposed method outperformed baseline CNN methods. The traditional CNN uses cross-entropy loss, which is limited to learn the distance between the true probability and the predicted one. On the other hand, contrastive loss calculates the distance between the embedding data, it learns to embedded close to each other samples from the same class and to be embedded in a distance larger than the margin samples from different classes [28-31].

In addition, the limitation of EEG studies is that there is a small number of data. If the number of data is too small, an approach using deep learning is not appropriate. However, it is difficult to obtain a large amount of data when using EEG signals due to the nature of the experiments, especially in imagined speech paradigms. There are only 40 trials per class in the used database [19]. In some studies, only 12 trials per class were recorded [32]. In this case, the deep learning method with cross-entropy does not learn well and tends to overfit. However, since the Siamese neural network using contrastive loss was originally proposed in one-shot learning, it can be used even for a small amount of data. Siamese neural network uses two pair samples as input. Therefore, it has the advantage of being able to increase the number of data by the number of combinations [30]. The study used as many combinations as possible to increase performance.

The layer also consists of three convolution layers, three max-pooling layers, and four fully-connected layers. The reason for using four fully-connected layers is that they act as a deep neural network, which learns more discriminant features. Our network performed better than when we applied contrastive loss to EEGNet [1] or DeepConvNet [25], which are often used in EEG analysis. Consequently, our network is more suitable for the imagined speech analysis.

As a result, our method showed an improvement of more than 9.93% over baseline methods. Moreover, statistical analysis showed significant differences between our proposed and other methods. The average performance across subjects was 31.40 ± 2.73%. Subject 4 showed the highest performance at 36.51 ± 4.13%. Therefore, we confirm that the Siamese neural network reduces high-dimensional data (specifically to 8 dimensions) while minimizing the loss of information. Additionally, our Siamese neural network classifies EEG signals from imagined speech without removing extra noise (i.e., removal of electrooculography). EEG signals have a low signal-to-noise ratio [28], which cannot be solved by preprocessing alone. However, our method seems to partially overcome it.

V. CONCLUSION

In this study, we investigated the end-to-end Siamese neural network encoder-based classification approach to classifying imagined speech using EEG. The data consisted of EEG for 6-class imagined speech (word). Our approach consisted of training the Siamese neural network using raw data and classify obtained embedding data.

Many previous studies aimed to improve the performance of imagined speech, however, most of them focus on preprocessing or feature extraction methods. Moreover, among deep learning studies, there are few approaches that use raw data as input to the network. We propose an approach to improve classification performance using raw data. In addition, we investigated a deep learning approach that can learn well without overfitting even when using a small number of trials.

In conclusion, we obtained an average accuracy of 31.40 ± 2.73% more than 9.93% higher than the baseline methods. The results showed that contrastive loss-based Siamese neural network has the potential to classify imagined speech. Siamese neural network is a deep metric learning-based approach. We decided to investigate other metric learning methods which we believe have the potential to increase performance. In addition, Siamese neural network acts as a feature extraction method, and *k*-NN learns to classify embedding vectors obtained through Siamese neural network. Our method has shown the possibility of using the imagined speech paradigm as an intuitive BMI for real-life environments.

ACKNOWLEDGMENT

The authors thank to J. Kalafatovich and D.-K. Han for useful discussion.